\documentclass{ws-procs975x65}            

\usepackage{amsmath}
\usepackage{slashed}

\newcommand{\be}{\begin{equation}}
\newcommand{\ee}{\end{equation}}
\newcommand{\bea}{\begin{eqnarray}}
\newcommand{\eea}{\end{eqnarray}}
\begin{document}
\title{Pair creation in electric fields, 
renormalization, and backreaction}

\author{Antonio Ferreiro$^*$, Jose Navarro-Salas$^{**}$, and Silvia Pla$^\dagger$}
\address{Departamento de Fisica Teorica and IFIC, Universidad de Valencia-CSIC,\\
Facultad de Fisica, Burjassot, 46100, Spain\\
$^*$E-mail: antonio.ferreiro@ific.uv.es\\
$^{**}$E-mail: jnavarro@ific.uv.es\\
$^{\dagger}$E-mail: silvia.pla@uv.es}

\begin{abstract}
We consider pair production  phenomena in spatially homogeneous strong electric fields.  We focus on spinor QED in two-dimensions  and discuss the potential ambiguity in the adiabatic order assignment  for the electromagnetic potential required to fix the renormalization subtractions. This ambiguity can be univocally fixed by imposing, at the semiclassical level,  stress-energy conservation when both electric and gravitational backgrounds are present.
\end{abstract}


\bodymatter

\section{Introduction}
A time-dependent gravitational field yields the creation operators of  quantum fields to evolve into a superposition of creation and annihilation operators. This produces the spontaneous creation of particle-antiparticle pairs out of the vacuum. 
This effect was first discovered,
including the precise probability distribution of the produced particles,
in the physical context of an expanding universe \cite{parker68}. A similar superposition of creation and annihilation operators takes place if the quantized field is coupled to a time-varying  gauge field background \cite{BI}. In this contribution we want to focus on the particle creation phenomena induced by electric fields in presence of gravity, which are of major relevance in  astrophysics and cosmology \cite{ruffini}.

A fundamental problem in the physical understanding of gravitational particle creation processes is the calculation of the expectation values $\langle T_{\mu\nu} \rangle$.
The computations require methods of regularization and renormalization to deal with new ultraviolet  divergences (UV) not present in Minkowski space.  Equivalently, in a time-dependent electric field the fundamental problem is to evaluate the renormalized electric current $\langle j_\mu \rangle$, which acts as the proper source of the semiclassical Maxwell equations.

In this work we discuss an improvement of the adiabatic regularization method, originally introduced in cosmological scenarios and for quantized scalar fields \cite{parker-fulling}, to  include homogeneous electric fields and Dirac quantized fields. 
We will reexamine the method to consistently deal with both electric and gravitational fields. By doing this we will fix an inherent ambiguity of the method. The adiabatic order assignment of the vector potential has been traditionally assumed in the literature of order zero. We will  argue that the correct adiabatic order assignment is one, instead of zero, at least if a gravitational field is present. This problem has been discussed for scalar fields in \cite{FN, FNP}. Here we extend the discussion to Dirac fields. To focus on the main ideas we will restrict the analysis to two-dimensional spacetime.

\section{Spinor QED$_2$ and the adiabatic renormalization scheme}

We  consider  two-dimensional spinor QED in an expanding spacetime described by the metric  $ds^2 = dt^2 - a^2(t)dx^2$. 
The classical action is given by
\bea
\mathcal{S}=\int dx^2 \sqrt{-g}\left(-\frac14 F_{\mu\nu}F^{\mu\nu}+i\bar{\psi}\slashed{D}\psi-m \bar{\psi}\psi\right) \ , 
\eea
and the corresponding Dirac equation  reads
\bea
 (i \underline{\gamma}^{\mu}D_{\mu}-m)\psi=0\label{diraceq} \  ,
 \eea
 where $D_{\mu} \equiv \partial_{\mu}-\Gamma_{\mu} -i q A_{\mu}$ and $\Gamma_\mu$ is the spin connection. $\underline{\gamma}^{\mu}(x)$ are the spacetime-dependent Dirac matrices satisfying the anticommutation relations $\{\underline{\gamma}^{\mu},\underline{\gamma}^{\nu}\}=2g^{\mu\nu}$. These  gamma matrices are related with the Minkowskian ones by $\underline{\gamma}^0 (t) = \gamma^0$ and $\underline{\gamma}^1 (t) = \gamma^1 / a(t)$, and the components of the spin connections are $\Gamma_0=0$ and $\Gamma_1=(\dot{a}/2) \gamma_0 \gamma_1$. Therefore,  $\underline{\gamma}^{\mu}\Gamma_{\mu}=-\frac{\dot{a}}{2a}\gamma_0$. 
It is very convenient to fix the gauge for the vector potential as $A_{\mu}=(0,-A(t))$. The Dirac equation \eqref{diraceq} becomes\cite{FN, BFNV}
  \bea
 \label{Dirac}\left(i \gamma^0 \partial_0 +\frac{i}{2}\frac{\dot{a}}{a}\gamma^0+\left(\frac{i}{a}\partial_1+\frac{q A_1}{a}\right)\gamma^1-m\right)\psi=0.
 \eea
From now on we will use  the Weyl representation (with $\gamma^5 \equiv \gamma^0\gamma^1$)
\bea
\gamma^0 = \scriptsize
\left( {\begin{array}{cc}
 0 & 1  \\
 1& 0  \\
 \end{array} } \right),\hspace{2cm} 
\gamma^1 = \scriptsize \left( {\begin{array}{cc}
 0 & 1  \\
 -1& 0  \\
 \end{array} } \right), \hspace{2cm} \gamma^5 = \scriptsize \left( {\begin{array}{cc}
 -1 & 0  \\
 0& 1 \\
 \end{array} } \right) \nonumber
 \ . \eea
 We expand the quantized field in  momentum modes 
\bea
\label{spinorbd}\psi(t, x)=\int_{-\infty}^{\infty} dk \left[B_k u_k(t, x)+D^{\dagger}_k v_k(t, x)\right] \ , 
\eea
where the two independent spinor solutions are
\bea
 u_{k}(t, x)=\frac{e^{ikx}}{\sqrt{2\pi a}} \scriptsize \left( {\begin{array}{c}
 h^{I}_k(t)   \\
 -h^{II}_k (t) \\
 \end{array} }\right), \hspace{1.5cm} v_{k}(t, x)=\frac{e^{-ikx}}{\sqrt{2\pi a}} \scriptsize \left( {\begin{array}{c}
 h^{II*}_{-k} (t)  \\
 h^{I*}_{-k}(t)  \\
 \end{array} } \right)
 \label{spinorde}
\ , \eea
and $B_k$ and $D_k$ the creation and annihilation operators which fulfill the usual anti-commutation relations. The field equations (\ref{Dirac}) are now converted into 
\bea \label{system}
&&\dot{h}^{I}_k-\frac{i}{a} \left(k+qA\right)h^{I}_k-i m h^{II}_k=0\nonumber \\
&&\dot{h}^{II}_k+\frac{i}{a}\left(k+qA\right)h^{II}_k-i m h^{I}_k=0 \ . 
\eea
We also assume  the normalization condition $ |h_k^{I}|^2+|h_k^{II}|^2=1$.
The classical electric current is given by 
$j^{\nu}=-q \bar \psi\underline{\gamma}^{\nu} \psi $, 
and the formal expression for the vacuum expectation value of $\langle j^x \rangle$ is 
	\bea
	\langle j^x \rangle&&=\frac{q}{2\pi a^2} \int_{-\infty}^{\infty}  dk \left(\left|h_k^{II}\right|^2-\left|h_k^{I}\right|^2\right) \label{corriente1} \ .
	\eea
The above expression possesses, as expected,  UV divergences. To obtain the finite, physical values, we have to perform appropriate subtractions
\be 	\langle j^x \rangle_{ren}=\frac{q}{2\pi a^2} \int dk \left( |h_k^{II}|^2-|h_k^{I}|^2 - SUBSTRACTIONS\right) \ . \ee 
The semiclassical Maxwell equations for $F_{\mu\nu}= \partial_\mu A_\nu - \partial_\nu A_\mu$ are 
\bea
\nabla_{\mu}F^{\mu\nu}=\langle j^{\nu} \rangle_{ren} \label{semimax}  \ . \eea
Equations \eqref{system} and \eqref{semimax} determine the continuous interchange of energy between the electric field and matter, via charged pair production and backreaction.
Now, the main problem is to obtain the required subtractions consistently. In this context, the most natural way to determine the renormalization subtractions is the adiabatic regularization method.
The basic principles for scalar fields can be borrowed from Ref. \cite{parker-toms}, while for spin-$1/2$ fields one can see \cite{rio1, BFNV}. 
The main idea here is to consider an adiabatic expansion of the mode functions. For scalar fields, this expansion is based on the WKB-type ansatz , namely
\be h_{k}(t) =\frac{1}{\sqrt{\Omega_k(t)}}e^{-i \int^{t}\Omega_k(t')dt'} \ , \hspace{0.5cm}  \Omega_k(t) = \omega^{(0)}_k + \omega_k^{(1)} + \omega_k^{(2)} + \cdots \, \ee
where the order of the expansion is determined by  the number of  derivatives of the background fields. A different type of expansion is required for fermionic fields\cite{rio1, BFNV}. A very crucial point to properly define the adiabatic expansion, for both scalar and Dirac fields, is to fix the leading order. The  zeroth order adiabatic term is determined by $ \omega^{(0)}_k$. 
\section{Two attempts toward the backreaction equations in Minkowski space}
Let us first assume for simplicity that we are in Minkowski space and $a(t)=1$. Therefore, the gauge field is now the only background field. 
In this scenario it is very natural to define $\omega^{(0)}_k = \sqrt{(k+qA)^2 + m^2} $.
 This means that the adiabatic order assignment for the gauge field $A(t)$ has been implicitly chosen as $0$, as  first assumed in Refs. \cite{Cooper1} and in all subsequent papers on this topic. Therefore, $\dot A(t)$ should be of order $1$, etc. 
The proposed renormalized current is then given by
\bea \langle j^{x}\rangle_{ren}^{I}&=&q \int_{-\infty}^{\infty} \frac{dk}{2 \pi} \left (|h_k^{II}|^2-|h_k^{I}|^2 -\frac{(k+qA)}{\omega^{(0)}_k}  \right )
\label{jren0}\eea
Plugging this expression into the semiclassical Maxwell equations \eqref{semimax}, we get the following backreaction equation,
\be -\dot{E}= \ddot{A}=q \int_{-\infty}^{\infty} \frac{dk}{2 \pi} \left [|h_k^{II}|^2-|h_k^{I}|^2-\frac{\left(k+q A\right)}{\sqrt{(k+qA)^2 +m^2}} \right ],\label{bkmax}\ee
together with the equation for the field modes (\ref{system}). It is important to stress that the above semiclassical Maxwell equations are compatible with the conservation of the energy if the renormalized stress-energy tensor is constructed by subtracting up to the zero adiabatic order 
\bea
\langle \rho \rangle_{ren}^{I}=\langle T_{00}\rangle_{ren}^{I}=\frac{1}{2\pi }\int_{-\infty}^{\infty} dk i \left [h_k^{II}\dot{h}_k^{II*}+h_k^{I}\dot{h}_k^{I*}\right]+\omega_k^{(0)} \label{fden} \ . \eea
It is easy to check that $ \partial_\mu \langle T^{\mu\nu} \rangle_{ren} + \partial_\mu T^{\mu\nu}_{elec} =0$, where $T_{\mu\nu}^{elec} =   \frac12 E^2 \eta_{\mu\nu}$.


Alternatively, we can define 
\bea
 \label{2omega}\omega^{(0)}_k = \sqrt{k^2 + m^2} \equiv \omega ,
 \eea
which assumes that  $A$ is of adiabatic order $1$, and proceed according to the rules of the adiabatic expansion. With this choice we obtain
\bea
\langle j^x \rangle_{ren}^{II}=q \int_{-\infty}^{\infty}  \frac{dk}{2\pi } \left( |h_k^{II}|^2-|h_k^{I}|^2-\frac{k}{\omega}-\frac{q m^2} {\omega^3}A\right)\label{jren1} \ .
\eea
Hence, the semiclassical Maxwell equation reads $-\dot{E}= \ddot{A}=\langle j^x \rangle_{ren}^{II}$, and the mode functions $h_k^{I}$ and $h_k^{II}$ satisfy again eq. \eqref{system}.
%
The choice \eqref{2omega} is also compatible with the conservation of energy, defined now by subtracting up to second adiabatic order
\bea
\langle T_{00}\rangle_{ren}^{II}=\frac{1}{2\pi }\int dk i \left [h_k^{II}\dot{h}_k^{II*}+h_k^{I}\dot{h}_k^{I*}\right]+\omega +\frac{k q A}{\omega}+\frac{m^2 q^2 A^2}{2\omega^3} \ . \label{fden}  \eea

It is immediate to see that the fermionic currents \eqref{jren0} and \eqref{jren1} are equivalent, 
\bea
 \bigtriangleup \langle j^{x}\rangle_{ren}= q \int_{-\infty}^{-\infty} \frac{dk}{2 \pi } \left [\frac{k}{\omega} - \frac{\left(k+q A\right)}{\sqrt{(k+qA)^2 +m^2}}+\frac{m^2 q A}{\omega^3}  \right ] =0 \label{difference}\ .
\eea
The first two terms of the equation above correspond to linearly divergent integrals, differing by a constant shift, and hence, their difference is finite $-q^2A(t)/\pi$. On the other hand, the last term is a finite integral, which cancels out with the previous quantity.  
In conclusion, in Minkowski space the  choice of the adiabatic order of the background field $ A (t) $ does not affect the physical observables and, therefore, both options are equally valid. However, this conclusion is misleading as soon as one introduces the gravitation field.

\section{The role of gravity}

Now, let us assume that our space-time is a two-dimensional expanding universe with metric $ds^2= dt^2 - a^2(t) dx^2$. In this case, the prescription for $a(t)$ is to fix it at adiabatic order 0, and the rules for the adiabatic subtraction terms are univocally fixed according to the scaling dimension of the relevant operators. The stress-energy tensor must be renormalized at second adiabatic order \cite{birrell-davies}, while the electric current should be renormalized at adiabatic order $1$. 
With this restrictions, the only possibility is to choose $A(t)$ of adiabatic order $1$, in the same footing as $\dot a(t)$. 
That is, we should have a hierarchy between the two background fields.  The leading order corresponds to gravity, and consequently one should replace the definition (\ref{2omega}) by 
\be \label{3omega}\omega^{(0)}_k = \sqrt{k^2/a^2+ m^2}\equiv \omega \ , \ee
instead of the naive generalization $\omega^{(0)}_k = \sqrt{(k+qA)^2/a^2  + m^2} \ $. This point has been overlooked in the literature, as recently stressed in Ref. \cite{FN}.
 The gauge field should enter at the next to leading order in the adiabatic expansion.

Therefore, the renormalized expression for the electric current should be
\bea
\langle j^x \rangle_{ren}=\frac{q}{2\pi a^2} \int_{-\infty}^{\infty}  dk \left( |h_k^{II}|^2-|h_k^{I}|^2-\frac{k}{a\omega}-\frac{q m^2} {a\omega^3}A\right)\label{jren} \ ,
\eea
and the backreaction equations are 
\bea &&-\dot{E}=\frac{\ddot A}{a}-\frac{\dot A}{a}\frac{\dot{a}}{a}=\frac{q}{2\pi a} \int_{-\infty}^{\infty} dk \left( |h_k^{II}|^2-|h_k^{I}|^2  -\frac{k}{a\omega}-\frac{q m^2}{a\omega^3}A\right) \ \label{bkmax} \ , \eea
together with the equation for the field modes \eqref{system}.

As in Minkowski space,  one should expect stress-energy conservation, namely $\nabla_\mu \langle T^{\mu\nu} \rangle_{ren} + \nabla_\mu T^{\mu\nu}_{elec} =0$, with $T_{\mu\nu}^{elec} =   \frac12 E^2 g_{\mu\nu}$. After the required adiabatic subtractions, one obtain the following expressions for the diagonal components of $\langle T_{\mu\nu} \rangle$:
 {\small\bea
\langle T_{00}\rangle_{ren}=\frac{1}{2\pi a}\int_{-\infty}^{\infty}  dk i \left [h_k^{II}\dot{h}_k^{II*}+h_k^{I}\dot{h}_k^{I*}\right]+\omega +\frac{k q A}{\omega}+\frac{m^2 q^2 A^2}{2a^2\omega^3}-\frac{k^2m^2 \dot a^2}{8 a^4 \omega^5},\label{scden}\eea}
 {\small \bea
\langle T_{11} \rangle_{ren}=&&\frac{1}{2\pi}\int_{-\infty}^{\infty}  dk (k+qA(t))\left(|h_k^I|^2-|h_k^{II}|^2\right)+\frac{k^2}{a \omega}+\frac{ k m^2 qA}{a \omega ^3}+\frac{k qA}{a \omega }-\frac{m^4 \ddot a}{4 \omega ^5}\nonumber \\&&+\frac{m^2 \ddot a}{4 \omega ^3}+\frac{5 m^6 \dot a^2}{8 a \omega ^7}-\frac{3 m^4 \dot a^2}{4 a \omega ^5}+\frac{m^2 \dot a^2}{8 a
   \omega ^3}+\frac{3 m^4 q^2A^2}{2 a \omega ^5}-\frac{ m^2 q^2A^2}{2 a \omega ^3} \ . \label{fT11}
\eea}
The conservation equation for the zeroth component can be decomposed as
{\small \bea
\nabla_{\mu}\langle T^{\mu0}\rangle_{ren}+ \nabla_\mu T^{\mu0}_{elec}= \partial_0\langle T_{00}\rangle_{ren}+\frac{\dot a}{a}\langle T_{00}\rangle_{ren}+\frac{\dot a}{a^3}\langle T_{11}\rangle_{ren}+ \partial_0 T_{00}^{elec}=0.\label{conservacion2} \
\eea}
Plugging  \eqref{scden} and  \eqref{fT11} into \eqref{conservacion2} and using the equations  \eqref{system} we get
 \bea
\nabla_{\mu}\langle T^{\mu0}\rangle_{ren}+ \nabla_\mu T^{\mu0}_{elec}=\frac{ \dot{A}}{a}\left(\frac{ \ddot{A}}{a}-\frac{ \dot{A}\dot{a}}{a^2}- \langle j^x \rangle_{ren} \right)=0 \label {conservation3} \ .
\eea
Note that the factor in parentheses is precisely 
the semiclassical Maxwell equation for the electric field \eqref{bkmax}. A similar result is trivially obtained for 
the remaining component: $\nabla_{\mu}\langle T^{\mu1}\rangle_{ren}+ \nabla_\mu T^{\mu1}_{elec}=0$.
 Therefore, the semiclassical Maxwell equations must be satisfied to ensure stress-energy conservation. 

If one assumes  that $A(t)$ is of adiabatic order $0$, the stress-energy conservation does not hold anymore. With this alternative adiabatic assignment, and renormalizing $\langle T^{\mu\nu}\rangle$ up to and including the second adiabatic order, one obtains 
\bea
\nabla_{\mu}\langle T^{\mu0}\rangle_{ren}+ \nabla_\mu T^{\mu0}_{elec}\neq 0 \ . \eea
In this case, the left-hand side in the above equation is proportional to  $E \langle j^x\rangle^{(2)}$, where $\langle j^x\rangle^{(2)}$ is the second adiabatic order of the electric current, which cannot be properly absorbed into the renormalization substractions of the electric current. We note that another inconsistency is the disagreement with the trace anomaly \cite{FN} in the massless limit.

\section{Conclusions}\label{conclusions}

The role of gravity is essential to fix the ambiguity in the adiabatic order assignment of the electromagnetic potential $A(t)$, and therefore to obtain the correct renormalization subtractions of  physical observables. Although in absence of a gravitational background the two different prescriptions for 
the adiabatic subtraction terms are equivalent, when gravity is incorporated into the game only the adiabatic order assignment $1$ for the gauge field $A(t)$ is compatible with stress-energy conservation.


\begin{thebibliography}{10}


%
%
%
%
%
%
%
%
%
%
%
%







\bibitem{parker68} L.~Parker, {\it Phys.~Rev.~Lett.} {\bf 21}, 562 (1968); {\it Phys.~Rev.~D} {\bf 183}, 1057 (1969).



\bibitem{BI} E. Brezin and C. Itzykson, {\it Phys. Rev. D} {\bf 2},  1191 (1970).

\bibitem{ruffini} 
  R.~Ruffini, G.~Vereshchagin and S.~S.~Xue, {\it Phys.\ Rept.}  {\bf 487}, 1 (2010).
  C. Stahl, E. Strobel and S.-S. Xue, {\it Phys. Rev. D} {\bf 93}, 025004 (2016). 

\bibitem{parker-fulling} L.~Parker and S.~A.~Fulling, {\it Phys.~Rev.~D} {\bf 9}, 341 (1974). 
S. A. Fulling and L. Parker, {\it Ann.~Phys.} (N.Y.) {\bf 87}, 176 (1974). S. A. Fulling, L. Parker and B. L. Hu, {\it Phys. Rev. D} {\bf 10}, 3905 (1974). 
T.~S.~Bunch, {\it J.~Phys.~A} {\bf13}, 1297 (1980). P.~R.~Anderson and L.~Parker, {\it Phys.~Rev.~D} {\bf 36}, 2963 (1987). I.~Agullo, J.~Navarro-Salas, G.~J.~Olmo and  L.~Parker, {\it Phys.~Rev.~Lett.} {\bf 103}, 061301 (2009). {\it Phys.~Rev.~D} {\bf 81}, 043514, (2010).  

\bibitem{FN} A. Ferreiro and J. Navarro-Salas, {\it Phys. Rev. D} {\bf 97}, 125012   (2018).


\bibitem{FNP} A. Ferreiro, J. Navarro-Salas and S. Pla, {\it Phys. Rev. D}{\bf 98}, 045015 (2018).


\bibitem{BFNV}  F. J. Barbero G., A. Ferreiro, J. Navarro-Salas and E. J. S. Villase\~nor, {\it Phys. Rev. D}{\bf 98}, 98, 025016 (2018).
\bibitem{parker-toms}L.~Parker and D.~J.~Toms, {\it Quantum Field Theory in Curved Spacetime: Quantized Fields and Gravity}, CUP, Cambridge, UK (2009).





\bibitem{rio1} A. Landete, J. Navarro-Salas and F. Torrenti, {\it Phys. Rev. D} {\bf 88}, 061501 (2013); {\it Phys. Rev. D} {\bf 89} 044030 (2014). A. del Rio, J. Navarro-Salas and F. Torrenti, {\it Phys. Rev. D} {\bf 90}, 084017 (2014). 
S.~Ghosh, {\it Phys.~Rev.~D} {\bf 91}, 124075 (2015); {\it Phys.~Rev.~D} {\bf 93},   044032 (2016). A.~del Rio and  J.~Navarro-Salas, {\it Phys.~Rev.~D} {\bf 91}, 064031 (2015). 
A.~del Rio, A. Ferreiro, J.~Navarro-Salas and F.~Torrenti, {\it Phys.~Rev.~D} {\bf 95}, 105003 (2017). 

\bibitem{Cooper1}F. Cooper and E. Mottola, {\it Phys. Rev. D} {\bf 40}, 456 (1989). Y. Kluger et al. {\it Phys. Rev. Lett.} {\bf 67}, 2427 (1991).


%




\bibitem{birrell-davies} N.~D.~Birrell  and P.~C.~W.~Davies, {\it Quantum Fields in Curved Space}, CUP, Cambridge, UK (1982).






















\end{thebibliography}
\end{document}